 \documentclass[preprint,aps,showpacs]{revtex4}
 \pdfoutput=1
\usepackage[pdftex]{graphicx}

\begin{document}
  \newcommand{\Qed}{\rule{2.5mm}{3mm}}
 \newcommand{\balpha}{\mbox{\boldmath {$\alpha$}}}
 \newcommand{\mali}[1]{{\scriptscriptstyle#1}}
 \newcommand{\NP}{\mali{NP}}
  \newcommand{\SP}{\mali{SP}}
 \def\Tr{{\rm Tr}}
 \def\(#1)#2{{\stackrel{#2}{(#1)}}}
 \def\[#1]#2{{\stackrel{#2}{[#1]}}}
 \def\A{{\cal A}}
 \def\B{{\cal B}}
 \def\Sb#1{_{\lower 1.5pt \hbox{$\scriptstyle#1$}}}

\title{
"An effective two dimensionality" cases bring a new hope to the Kaluza-Klein[like] theories 
}
\author{D. Lukman}
\author{ N. S. Manko\v c Bor\v stnik}
\address{ Department of Physics, FMF, University of
Ljubljana, Jadranska 19, 1000 Ljubljana}
\author{ H. B. Nielsen}
\address{Department of Physics, Niels Bohr Institute,
Blegdamsvej 17,\\
Copenhagen, DK-2100}

\begin{abstract} 
One step towards realistic Kaluza-Klein[like] theories and a loop hole through the 
Witten's "no-go theorem" is presented for cases which we call {\em an effective two dimensionality} cases:  
In $d=2$ the equations of motion following from the action with the linear curvature leave  
spin connections and zweibeins undetermined. 
We present the case of a spinor in $d=(1+5)$ compactified on a formally {\em infinite} disc with 
the {\em zweibein} which makes a disc curved on an almost $S^2$ and with the {\em spin connection} 
field  which allows on such a sphere only one massless {\em normalizable} 
spinor state of a particular charge, 
which couples the spinor chirally to the corresponding Kaluza-Klein gauge field. 
We assume no external gauge fields. 
The masslessness of a spinor is achieved by the choice   
of a spin connection field (which breaks left-right symmetry), the zweibein 
and the normalizability 
condition for spinor states, which guarantee  a discrete spectrum forming the complete basis. 
We discuss the meaning of the hole, which manifests the 
noncompactness of the space. 
\end{abstract}



\pacs{11.10.Kk, 11.25.Mj, 12.10.-g, 04.50.-h}

\maketitle

\date{\today}

\section{Introduction}
\label{introduction}

The idea of Kaluza and Klein~\cite{kk} of obtaining the electromagnetism 
- and under the influence of their idea  nowadays also the  weak and colour 
fields~\cite{geogla,chofre,zee,salstr,mec,dufnilspop,daess,wet} - 
from purely gravitational degrees of freedom connected with having extra dimensions 
is very elegant. More than twenty fives years ago the Kaluza-Klein[like] theories were 
studied very intensively 
by many authors~\cite{wet,dufnilspop,zelenaknjiga,horpal}. Although the breaking of the 
symmetry of the starting Lagrange density to the low energy effective ones 
(that is to the charges and correspondingly to the 
gauge fields assumed  by the {\it standard model of the electroweak and colour 
interactions}) seem very promising, 
the idea of Kaluza and Klein was almost killed by the "no-go theorem" of 
E. Witten~\cite{witten}  telling that 
these kinds of Kaluza-Klein[like] theories with the gravitational fields only (that is with 
vielbeins and spin connections) have  severe difficulties with obtaining  massless fermions 
chirally coupled to the Kaluza-Klein-type gauge fields in $d=1+3$, 
as required by the {\it standard model}.  
There were attempts to  escape from the "no-go theorem" in compact extra spaces 
by having torsion~\cite{salstr}, %
or by having an orbifold structure~\cite{orbifold}, or by putting extra gauge fields by hand in 
addition to gravity in higher dimensions~\cite{sap}, which is no longer the pure 
Kaluza-Klein[like] theory and loses accordingly the elegance. 

Since there is the assumption that the space is {\em compact} in the "no-go theorem" of 
E. Witten, there are also the attempts to 
achieve masslessness by appropriate choices of vielbeins in {\em noncompact} 
spaces, one of works~\cite{wet} is commented in the footnote~\footnote{The author of the ref.~\cite{wet} 
proposes, for example,  the "squashed" $S^2$ sphere, recognizing that with the zweibein  of $S^2$ 
(he calls in this case $S^2$ a compact space) there are  
no massless spinor states, while with at least a little "stronger"  zweibein than with that  
of $S^2$ (like with $f= (1+(\frac{\rho}{2\, \rho_0})^{2+k})$, with $0< k \le 2$,  
$k=0$ reproduces $S^2$) there are two massless states. 
Although 
the author wrote differently, these two 
massless states belong to the left and the right handed state with respect to $d=(1+3)$, and therefore 
not mass protected, and 
would correspondingly lead to massive fermion states
.}. 

There are several attempts to point out the importance of non compact extra dimensions, like 
~\cite{mmWeakScale98di}, many of them 
surveyed in~\cite{rubakovExtraDim}. These attempts do not really try to  keep the 
Kaluza-Klein approach in the original elegant version, they rather embed strings, membranes, 
p-branes into higher dimensional spaces. 
The most popular models of this kind are probably
Randall-Sundrum models~\cite{rs}.

We are interested in this paper in  extra dimensions in the Kaluza-Klein sense: that is as a 
possibility that the gravity (and only gravity)
in extra dimensions manifests as the {\it standard model} gauge fields in $(1+3)$, coupled to 
 the corresponding charges.
In refs.~\cite{hnkk06} we achieved masslessness of spinors in the pure Kaluza-Klein[like] theory 
(for the case of $M^{1+5}$ manifold broken into $M^{1+3} \times$ an infinite disc) with  the appropriate 
choice of a boundary limiting the extra dimensions on a finite surface on a disc. 

In the proposed paper we take the whole two dimensional plane, and  roll it up into  an almost 
$S^2$ with one point - the south pole - excluded. 
It is our choice of a  {\em zweibein} which forces the  
two extra dimensions into 
an almost 
$S^2$. Thus, although it has a finite volume (namely the surface of $S^2$), 
the space is {\em non compact}. We require   spinor states to be  in the fifth and sixth dimensions   
{\em normalizable}~\footnote{In the ref.~\cite{wet}, mentioned 
and discussed in the previous footnote, this idea of a finite volume of a noncompact space, as well as the 
normalizability of states is already stressed.}, proving that the normalizable solutions form a complete 
set.  
It is our choice of a {\em particular spin connection field}, with the  strengths within an 
interval, which allows only  
one  normalizable massless state of a particular handedness (with respect to $(1+3)$), breaking the parity 
symmetry.  

The {\it finite volume}   of an {\em infinite} disc, an appropriate {\em choice of the spin connection field} 
with the strength $F$ allowed to be within the whole interval $0< 2F \le 1$ and the 
{\em normalizability} requirement  make the mass spectrum of our Hermitean 
Hamiltonian in a noncompact space discrete, with only one massless state of particular charge chirally 
coupled to the Kaluza-Klein gauge field. It is the sign of $F$ which makes a choice of the handedness of a 
massless state,  breaking the parity symmetry. 
The usually expected problem with extra non compact dimensions  
having a continuous spectrum is {\it not present} in our model.

For a particular choice of the strength of the spin connection field 
we find the states and the  spectrum (the masses) analytically. This mass spectrum of states forms the 
complete set 
on our almost $S^2$. For the remaining values of the strength, for all of  
which only one massless solution of a particular handedness in $(1+3)$ exists, 
it is not difficult to find the  recursive formulas for  normalizable solutions and the masses. 
Accordingly in this 
two dimensional noncompact space, with the spin connections and vielbeins which both are 
a part of the gravitational gauge fields and with no presence of an (additional) external field, 
the "no-go theorem" of E. Witten is not valid.

We also characterize 
the "singularity" which the spinor solutions "feel" on our infinite disc with the zweibein of a $S^2$ sphere, 
when treating the disc as the almost $S^2$ sphere, that is the $S^2$ sphere with the hole on the 
southern pole, so that we  have almost $M^{(1+3)}\times S^2$ case, 
that it is almost a compact space.

Let us add: As it is not difficult to recognize, the two dimensional spaces are very special~\cite{mil,deser}.  
Namely, in dimensions higher than two, 
when we have no fermions present and only  the curvature in the first power in the Lagrange 
density, the spin connections are normally determined from the vielbein fields, and   
the torsion is zero. In the two dimensional spaces, the vielbeins do not determine 
the spin connection fields. 
In the present article we pay attention to cases, which we call {\em an effective 
two-dimensionality}, when the spin connections are not
fully determined by the vielbeins.

In the here proposed types of models there is the chance for having chirally mass protected fermions 
in a theory in which the chirally 
protecting effective four dimensional gauge fields are {\it true} Kaluza-Klein[like] fields,  
the degrees of which inherit from the higher dimensional gravitational ones.
We are thus hoping for a revival of true Kaluza-Klein[like] models as candidates for 
phenomenologically viable models!

One of us has been trying for long to develop the  {\it approach unifying spins and charges and 
predicting families} 
(N.S.M.B.)~\cite{norma92939495,holgernorma20023} so that 
spinors which carry in $d\ge 4$ nothing but two kinds of the spin (no charges), would manifest 
in $d=(1+3)$ all the properties assumed by the {\it standard model} and does accordingly 
share with the Kaluza-Klein[like] theories the problem of masslessness of the fermions 
before the electroweak like types of break. We present briefly  the ideas of the 
{\it approach} in the 
footnote~\footnote{ The 
{\it approach unifying spin and charges and predicting families}~\cite{norma92939495}  
proposes in $d=(1 + (d-1))$  a simple starting action for spinors  with the two kinds of the 
spin generators ($\gamma$ matrices): the Dirac one, which takes care of the spin and the charges, 
and the second one, 
anticommuting with the Dirac one, which generates families. For the explanation of the 
appearance of the two kinds of the spin generators we invite the reader to look at the 
refs.~\cite{norma92939495,holgernorma20023} and the references therein. 
A spinor couples in $d=1+13$ 
to the vielbeins and (through two kinds of the spin generators to) the 
spin connection fields.  
Appropriate breaks of the starting symmetry lead to the 
left handed quarks and leptons in $d=(1+3)$, which carry the weak charge while the 
right handed ones are weak chargeless. The {\it approach} is offering the answers to  
the questions about the origin of families of quarks and leptons, about the explicit 
values of their masses and mixing matrices (predicting the fourth family to be possibly 
seen at the LHC or at somewhat higher energies)  
as well as about the masses of the scalar  and the weak gauge fields, about the dark 
matter candidates, and about breaking the discrete symmetries. 
There are many possibilities in the {\it approach } for breaking 
the starting symmetries to those of the {\it standard model}. These problems were studied 
in some crude approximations in refs.~\cite{norma92939495} and are under consideration~\cite{proc2010}.}.

Let us point out that in odd dimensional spaces and in even dimensional spaces divisible 
by four there is no mass protection in the 
Kaluza-Klein[like] theories~\cite{wet,hnkk06}. 
The spaces therefore, for which we can have a hope that the Kaluza-Klein[like] theories lead to 
chirally protected fermions and accordingly to the effective theory of the {\it standard model 
of the electroweak and colour interactions}, have $2(2n+1)$ dimensions. 
And breaking symmetries in such spaces, if starting with one Weyl spinor, and accordingly with the 
mass protected case, should again lead to mass protected cases in accordance with the {\it standard model}.

\section{The action, equations of motion, solutions, proofs and comments}
\label{starting action}

We prove in this section that in $M^{1+3} \times $ an infinite disc with the particular zweibein 
and spin connection on the disc there exists only one massless  normalizable (on the disc) fermion state of 
only one handedness and of a particular charge. It is accordingly mass protected. 
We also present proofs that the Hamiltonian is Hermitean and the spectra of normalizable states 
correspondingly discrete. For a 
particular strength of the spin connection field we present the spectrum and states. 
We discuss  the properties of solutions for the strengths allowed by the normalizability requirement.


Let us first repeat the four assumptions, stressed already in the introduction. 
\begin{enumerate}
\item We assume $2(2n+1)$-dimensional space,  in our case $n=1$,  with only gravity,    
described by the action~\footnote{We have proven in ref.~\cite{hnkk06} 
that only in even dimensional spaces of  $d=2$ 
modulo $4$ dimensions (\textit{i.e.} in $d=2(2n+1),$ $n=0,1,2,\ldots$) spinors  (they are allowed 
to be in families) of one handedness and with no conserved 
 charges gain no  Majorana mass.} 
  \begin{eqnarray}
         {\cal S} = \alpha \int \; d^d x \, E  {\cal\,  R}\,.
  \label{action}
  \end{eqnarray}
The Riemann scalar ${\cal R} = {\cal R}_{abcd}\,\eta^{ac}\eta^{bd}$ is determined by the Riemann tensor 
${\cal R}_{abcd} = f^{\alpha}{}_{[a} f^{\beta}{}_{b]}(\omega_{cd \beta, \alpha} 
- \omega_{ce \alpha} \omega^{e}{}_{d \beta} )$, 
with vielbeins $f^{\alpha}{\!}_{a}$~\footnote{$f^{\alpha}{}_{a}$ are inverted 
vielbeins to 
$e^{a}{}_{\alpha}$ with the properties $e^a{}_{\alpha} f^{\alpha}{\!}_b = \delta^a{}_b,\; 
e^a{}_{\alpha} f^{\beta}{}_a = \delta^{\beta}_{\alpha} $. 
Latin indices  
$a,b,..,m,n,..,s,t,..$ denote a tangent space (a flat index),
while Greek indices $\alpha, \beta,..,\mu, \nu,.. \sigma,\tau ..$ denote an Einstein 
index (a curved index). Letters  from the beginning of both the alphabets
indicate a general index ($a,b,c,..$   and $\alpha, \beta, \gamma,.. $ ), 
from the middle of both the alphabets   
the observed dimensions $0,1,2,3$ ($m,n,..$ and $\mu,\nu,..$), indices from 
the bottom of the alphabets
indicate the compactified dimensions ($s,t,..$ and $\sigma,\tau,..$). 
We assume the signature $\eta^{ab} =
diag\{1,-1,-1,\ldots,-1\}$.
} 
and  the spin connections $\omega_{ab\alpha}$
(the gauge fields of $S^{ab}= \frac{i}{4}(\gamma^a \gamma^b - \gamma^b \gamma^a)$). 
$[a\,\,b]$ means that the antisymmetrization must be performed over the two indices $a$ and $b$,
$E$ is the determinant of the inverse zweibein $e^{s}{}_{\sigma}, \, e^{s}{}_{\sigma} f^{\sigma}{}_{t}
\delta^s{}_t,\; $ (Eq.(\ref{fzwei})). 
\item Space $M^{1+5}$ has the symmetry of $M^{1+3} \times $ an infinite disc with the zweibein
on the disc 
\begin{eqnarray}
e^{s}{}_{\sigma} = f^{-1}
\pmatrix{1  & 0 \cr
 0 & 1 \cr},
f^{\sigma}{}_{s} = f
\pmatrix{1 & 0 \cr
0 & 1 \cr}\,,
\label{fzwei}
\end{eqnarray}
with 
\begin{eqnarray}
\label{f}
f &=& 1+ (\frac{\rho}{2 \rho_0})^2, 
\nonumber\\ 
x^{(5)} &=& \rho \,\cos \phi,\quad  x^{(6)} = \rho \,\sin \phi, \quad E= f^{-2}\,.\nonumber
\end{eqnarray}
The last relation follows  from $ds^2= 
e_{s \sigma}e^{s}{}_{\tau} dx^{\sigma} dx^{\tau}= f^{-2}(d\rho^{2} + \rho^2 d\phi^{2})$.
We use indices $s,t=5,6$ to describe the flat index in the space of an infinite plane, and 
$\sigma, \tau = (5), (6), $ to describe the Einstein index.  
$\phi$ determines the angle of rotations around  the axis perpendicular to the disc.
\item 
The spin connection field is chosen to be
\begin{eqnarray}
  f^{\sigma}{}_{s'}\, \omega_{st \sigma} &=& i F\, f \, \varepsilon_{st}\; 
  \frac{e_{s' \sigma} x^{\sigma}}{(\rho_0)^2}\, , \quad 
 0 <2F \le 1\, 
  ,\quad s=5,6,\,\,\; \sigma=(5),(6)\,. 
\label{omegas}
\end{eqnarray}
\item We require  normalizability of states $\psi$ on the disc
\begin{eqnarray}
\label{normalizability}
\int_{0}^{2\pi}\,d \phi \; \int_{0}^{\infty}\;E\,\rho d\rho \psi^{\dagger} \psi < \infty\,,   
\end{eqnarray}
as usual in quantum mechanics, allowing at most the plane waves normalized to the 
delta function: 
$\int_{-\infty}^{\infty}\,d x^{(5)} \,\int_{-\infty}^{\infty}\,d x^{(6)}\;E\, e^{i \vec{k}(\vec{x}-
\vec{x'})} = \delta^{2}(\vec{x}-\vec{x'})\,.$   
\end{enumerate}

Let us make now several statements, proofs of these statements and comments, which will help 
to clarify the meaning of the assumptions.\\ 
%
{\it Statement 1.:}
{\it In the absence of the fermion fields in $d=2$ any zweibein and any spin connection fulfills 
the equations of motion.}\\ 
{\it Proof 1.: }  The action of Eq.~(\ref{action})
 leads to the equations of motion~\cite{mil,norma92939495}  
\begin{equation}
\label{omegabcc}
(d-2)\,\omega_b{}^c{}_c = 
\frac{e^a{}_\alpha}{E} \partial_\beta\left(Ef^\alpha{}_{[a}f^\beta{}_{b]} \right), 
\end{equation}
which clearly demonstrate that  any spin connection  $\omega_b{}^c{}_c = 
 \omega_b{}^c{}_{\alpha} \, f^{\alpha}{}_{c}$ (which can in $d=2$ have  only two 
 different indices)  satisfies 
this equation. \\ 
{\it Comment 1.:}
For $d=2$ the variation of the action~(\ref{action})
  with respect to vielbeins leads to the equation
$-e_{s \, \sigma} R + 4 f^{\tau t} \omega_{st \sigma,\tau} = 0,$ 
which is 
zero for any $R$ ($-2 R + 2 R=0$). \\
{\it Statement 2.:} 
{\it The volume of this noncompact space} (which looks  almost as $S^2$ sphere) {\it  is finite.}  \\ 
{\it Proof 2.:} The volume is $\displaystyle \int^{\infty}_0 \, f^{-2}\rho \,d\rho= 
\pi\, (2 \rho_0)^2$.\\
{\it Comments 2.: } 
{\bf i.)} Finite volume helps to assure the existence of normalizable spinor states  on this disc. 
{\bf ii.)} The symmetry of this disc, which is the symmetry of $U(1)$ group, determines the charge of 
spinors in $d=(1+3)$.\\ 
{\it Statement 3.:} 
{\it The choice that }$M^{1+5}$ {\it breaks into }$M^{1+3} \times $ {\it an infinite disc 
with no gravity in }
$M^{1+3}$ {\it and with the zweibein of} Eq.~(\ref{fzwei}) {\it and the spin connection 
of} Eq.~(\ref{omegas}) {\it on an infinite disc makes the Lagrange density for a Weyl spinor} 
${\cal L}_{W} = \frac{1}{2} [(\psi^{\dagger} E \gamma^0 \gamma^a p_{0a} \psi) + 
(\psi^{\dagger} E \gamma^0\gamma^a p_{0 a}
\psi)^{\dagger}]$ {\it to be}
\begin{eqnarray}
 {\cal L}_{W} &=& \psi^{\dagger}\{E\gamma^0 \gamma^n p_n + E f \gamma^0 \gamma^s \delta^{\sigma}_s  ( p_{0\sigma} 
+  \frac{1}{2 E f}
\{p_{\sigma}, E f\}_- )\}\psi,\; n=0,1,2,3, \nonumber\\
&& p_{0\sigma} = p_{\sigma}- \frac{1}{2} S^{st}\omega_{st \sigma},
\label{weylE}
\end{eqnarray}
{\it with} $ E = \det(e^a{\!}_{\alpha}) = f^{-2}$, $f$ {\it is from } Eq.~(\ref{f}), 
{\it and with} $ \omega_{st \sigma}$ {\it from }Eq.~(\ref{omegas})~\footnote{One finds that 
 $ \omega_{cda} = \Re e \;\omega_{cda}, \;\; {\rm if \;\; c,d,a\;\; all\;\; different} $ while 
  $ \omega_{cda}= i\,\Im m\; \omega_{cda},\;\; \rm{otherwise}.$}.\\
{\it Proof 3.:}  Eq.~(\ref{weylE}) follows from the starting Lagrangean for a Weyl spinor interacting 
with only the vielbeins and spin connections  straightforwardly. \\
{\it Comment 3.:}
The Lagrange density  of Eq.~(\ref{weylE}) assures that the Hamiltonian is Hermitean. \\
{\it Statement 4.:} 
{\it Normalizability  condition for spinors on an infinite disc curled into an 
almost $S^2$ and with the spin connection of particular choice makes a choice of a 
spectrum which forms a complete set.} \\
%
{\it Proof 4.:} 
The Lagrange density of Eq.~(\ref{weylE}) leads to equations of motion~%
(Eqs.(\ref{weylEp},\ref{weylErho}, \ref{equationm56gen1}))
\begin{eqnarray}
\label{weylErho0}
if \, \{ e^{i \phi 2S^{56}}\, [\frac{\partial}{\partial \rho} + \frac{i\, 2 S^{56}}{\rho} \, 
(\frac{\partial}{\partial \phi}) -  \frac{1}{2 \,f} \, \frac{\partial f}{\partial \rho }\, 
(1- 2F \, 2S^{56})\,]
\, \} \, \psi^{(6)}
+ \gamma^0 \gamma^5 \, m \, \psi^{(6)}=0\,,
\end{eqnarray}
which look  for $F=1/2$ like Legendre equations~(Eq.~(\ref{equationm56u})).  
It is the sign of $F$ which makes a choice of the handedness of a massless state and breaks accordingly 
the parity symmetry. One can prove that 
the only normalizable eigenstates in the interval $0 \le \rho \le \infty$ are those with 
integer parameters $l$ and $n$, $(m\rho_0)^2= l(l+1)$, in Eqs.~(\ref{equationm56ux}). 
These states are Legendre polynomials and 
form the {\em complete set}. Solutions for  a non integer $n$  are  singular at 
$\rho =0$, while solutions with a non integer $l$ are singular at $\rho = \infty$, both 
singularities make the corresponding eigenstates not normalizable. \\
{\it Comments 4.:}
{\bf i.)} In the subsection \ref{equations} of this section the solutions of Eq.~(\ref{weylErho0}) are discussed
 for any choice of $F$ in the interval $0 <2F \le 1$. All the normalizable solutions can for any $F$ in this 
 interval be expressed as a normalizable superposition of a complete set of Legendre polynomials 
 and have the discrete spectrum. 
{\bf ii.)} In the limit when $\rho_0 \to \infty$, $f$ (in Eq.~(\ref{equationm56gen1}, next section)  goes to one 
and the two equations,  Eq.~(\ref{equationm56gen1}), define the recurrence relations 
between the Bessel functions of  an integer order ($\A_{n}(\rho m)= J_{n}(\rho m)$ 
and $\B_{n+1}(\rho m) = J_{n+1}(\rho m)$) for any mass $m$. 
Making the limit  $\rho_0 \to \infty$ in Eq.~(\ref{equationm56u}) in next section, with the discrete mass term 
$(m\rho_{0})^2= l(l+1)$  one again reproduces the Bessel equation, if putting  $l= m \rho_0$.  
(Bessels functions can be squared normalized only within a finite radius, determined by zeros.)
With $\rho_0$ going to infinity the distance between $m$-values solving
this constraint goes to zero, so that in this limit the system of allowed $m$ values approaches  
the continuum (all $m$ values). This is satisfactory because this limit corresponds to our 
already non-compact space approaches,
the usual flat two-dimensional space (with which one would have a truly
fully 5 +1 dimensional world in which of course the spectrum seen as $(3 +1)$-dimensional one 
should be continuous). 
{\bf iii.)}
For any finite $\rho_{0}$ can the plane wave in the fifth and sixth dimension be
expressed in terms of the Legendre polynomials. To a plane wave in general many Legendre
polynomials contribute, each corresponding to  a different mass. There is the solution 
for $2F=1$ which is independent of $x^{\sigma}\,, \, \sigma \in \{(5),(6)\}$. It 
corresponds to massless solution. This solution can be called the plane wave with
momentum zero.  
In the limit $\rho_0 \to \infty$ the definition for the plane waves in flat 
space follows. 
{\it Statement 5.:} 
{\it The zweibein} (Eq.(\ref{fzwei})) {\it and the spin connection} (Eq.(\ref{omegas}))  {\it with
the parameter} $F$  {\it within the interval} $0 <2F \le 1$ {\it allow 
only one massless spinor of a particular charge.}\\ 
{\it Proof 5.:} 
It is proven in the next subsection, in the last paragraph before  Eq.~(\ref{weylEp}), that it is 
the term $\psi^{\dagger}\, E f \gamma^0 \gamma^s \delta^{\sigma}_s  ( p_{0\sigma} +  \frac{1}{2 E f}
\{p_{\sigma}, E f\}_- )\psi$ in the Lagrange density~(Eq.(\ref{weylE})), which manifests as the mass 
term  $m$ in Eq.~(\ref{weylErho0}). There is a term in Eq.~(\ref{weylErho0}), namely 
$- if \,  e^{i \phi 2S^{56}} \,\frac{1}{2 \, f} \, \frac{\partial f}{\partial \rho }\, 
(1- 2F \, 2S^{56})\, \, \psi^{(6)}$, which clearly distinguishes between the two possible values of  
the spin operator $S^{56}$ in $d=5,6$, when this term applies on the state $\psi^{(6)}$, distinguishing 
 correspondingly also between  the two possible handedness of the state $\psi^{(6)}$ in 
$d=(1+3)$. It is shown in the next subsection that 
a normalizable  massless state ($m=0$ in Eq.~(\ref{weylErho0})) must fulfil  the condition:  
$( \; 0 \le (1-2F\,2 S^{56}) < 1)\;\psi^{(6)}$. The sign of $F$  chooses  the handedness 
of a massless normalizable spinor state.\\
{\it Comments 5.} 
{\bf i.)} Having the rotational symmetry around the axis perpendicular to the plane of the fifth and the sixth 
dimension it is meaningful to require that $\psi^{(6)}$ is the eigen function of the total angular momentum
operator $(M^{56}= x^5 p^6-x^6 p^5  + S^{56})$ in the fifth and sixth dimension $M^{56}= 
(-i \frac{\partial}{\partial \phi} + S^{56})\,;$
$M^{56}\,\psi^{(6)}= (n+\frac{1}{2})\,\psi^{(6)}$~(Eqs.(\ref{mabx},\ref{mabpsi}, \ref{weylErho})).
{\bf ii.)} The only massless state, which fulfills the normalization  
condition (see Eq.(\ref{masslesseqsolf})) for a positive $F$, is  the state with 
the property $2 S^{56}\,\psi^{(6)} = \psi^{(6)}$. Its charge (spin on the disc) is 
for $0 <2F \le 1$ equal to $\frac{1}{2}$ as it is shown in section~\ref{properties1+3}.
{\bf iii.)} All the other states are massive.   
{\bf iv.)} The current in 
the radial direction is for all these cases equal to zero for any $F$.

Detailed derivations of equations of motion and solutions are presented in subsection~\ref{equations} 
of this section. 

Let us summarize this section.
We have a Weyl spinor in $d=(1+5)$-dimensional space. This space breaks into 
$M^{1+3}$ cross an infinite 
disc with the zweibein which formally looks almost -- up to a hole in the southern pole -- as a $S^2$ sphere,
while a chosen  spin connection allows on such an infinite disc only 
one normalizable massless  state. The {\it Hamiltonian is Hermitean},   
the mass spectrum of {\it normalizable} states is correspondingly discrete and the 
probability for a fermion to escape out of the disc 
is zero~\footnote{It is expected that the zweibein  curving the infinite disc into an (almost $S^2$) and 
the spin connection, which breaks the parity symmetry and takes a part in determining equations of motion, 
appear dynamically, causing the  {\em "phase transition"}. Accordingly could dynamical fields by 
causing the phase transition 
restore the symmetry of $M^{1+5}$}. 

Allowing the whole interval of the strength of the spin connection fields ($0 < 2F \le 1$) 
the spin connection field 
is not  fine  tuned. 
For a particular choice of the constant of the spin connection field, that is for $2F=1$, 
the normalizable solutions are expressible with the Legendre polynomials and  
the massive states manifest a spectrum 
$m \rho_0 =l(l+1)$, with $l=0,1,2,\cdots$ and $-l \le n \le 1$.  $n+1/2$ is the charge of the spectrum.

A free choice of a zweibein and a spin connection field in the action of Eq.~(\ref{action}) 
is possible only in $d=2$ dimensional spaces (the presence of fermions might make this possible also for 
$d>2$).

{\it Let us point out that the "two dimensionality" can be simulated in any dimension larger than two,
if vielbeins and spin connections are completely flat in all but two dimensions} (this point is discussed 
also in the ref.~\cite{wet}).

\subsection{Solutions of the equations of motion  for spinors}
\label{equations}

We look for the  solutions  of the equations of motion (\ref{weylE}) for a spinor   
in $(1+5)$-dimensional space, which breaks into  
$M^{(1+3)} \times$  an infinite disc curved into a noncompact "almost" $S^2$ sphere as a superposition
of all  four ($2^{6/2 -1}$) states of a single Weyl representation. (We kindly ask the 
reader to see the technical details  about how to write 
a Weyl representation 
in terms of the Clifford algebra objects after making a choice of the Cartan subalgebra,  
for which we take: $S^{03}, S^{12}, S^{56}$ in the refs.~\cite{holgernorma20023}.)
In our technique 
one spinor representation---the four 
states, which all are the eigenstates of the chosen Cartan subalgebra with the eigenvalues $\frac{k}{2}$, 
correspondingly---are  
the following four products of projectors $\stackrel{ab}{[k]}$ and nilpotents 
$\stackrel{ab}{(k)}$: 
\begin{eqnarray}
\varphi^{1}_{1} &=& \stackrel{56}{(+)} \stackrel{03}{(+i)} \stackrel{12}{(+)}\psi_0,\nonumber\\
\varphi^{1}_{2} &=&\stackrel{56}{(+)}  \stackrel{03}{[-i]} \stackrel{12}{[-]}\psi_0,\nonumber\\
\varphi^{2}_{1} &=&\stackrel{56}{[-]}  \stackrel{03}{[-i]} \stackrel{12}{(+)}\psi_0,\nonumber\\
\varphi^{2}_{2} &=&\stackrel{56}{[-]} \stackrel{03}{(+i)} \stackrel{12}{[-]}\psi_0,
\label{weylrep}
\end{eqnarray}
where  $\psi_0$ is a vacuum state for the spinor state.
If we write the operators of handedness in $d=(1+5)$ as $\Gamma^{(1+5)} = \gamma^0 \gamma^1 
\gamma^2 \gamma^3 \gamma^5 \gamma^6$ ($= 2^3 i S^{03} S^{12} S^{56}$), in $d=(1+3)$ 
as $\Gamma^{(1+3)}= -i\gamma^0\gamma^1\gamma^2\gamma^3 $ ($= 2^2 i S^{03} S^{12}$) 
and in the two dimensional space as $\Gamma^{(2)} = i\gamma^5 \gamma^6$ 
($= 2 S^{56}$), we find that all four states are left handed with respect to 
$\Gamma^{(1+5)}$, with the eigenvalue $-1$, the first two states are right handed and the second two 
 states are left handed with respect to 
$\Gamma^{(2)}$, with  the eigenvalues $1$ and $-1$, respectively, while the first two are 
left handed 
and the second two right handed with respect to $\Gamma^{(1+3)}$ with the eigenvalues $-1$ and $1$, 
respectively. 
Taking into account Eq.~(\ref{weylrep}) we may write 
the most general wave function  
$\psi^{(6)}$ obeying Eq.~(\ref{weylErho0}) in $d=(1+5)$ as
\begin{eqnarray}
\psi^{(6)} = \A \,{\stackrel{56}{(+)}}\,\psi^{(4)}_{(+)} + 
\B \,{\stackrel{56}{[-]}}\, \psi^{(4)}_{(-)}, 
\label{psi6}
\end{eqnarray}
where $\A$ and $\B$ depend on $x^{\sigma}$, while $\psi^{(4)}_{(+)}$ 
and $\psi^{(4)}_{(-)}$  determine the spin 
and the coordinate dependent parts of the wave function $\psi^{(6)}$ in $d=(1+3)$ 
\begin{eqnarray}
\psi^{(4)}_{(+)} &=& \alpha_+ \; {\stackrel{03}{(+i)}}\, {\stackrel{12}{(+)}} + 
\beta_+ \; {\stackrel{03}{[-i]}}\, {\stackrel{12}{[-]}}, \nonumber\\ 
\psi^{(4)}_{(-)}&=& \alpha_- \; {\stackrel{03}{[-i]}}\, {\stackrel{12}{(+)}} + 
\beta_- \; {\stackrel{03}{(+i)}}\, {\stackrel{12}{[-]}}. 
\label{psi4}
\end{eqnarray}
Using $\psi^{(6)}$ in Eq.~(\ref{weylErho0}) and separating dynamics in $(1+3)$ and on the infinite disc 
the following relations follow, from which we recognize the mass term $m$:  
$\frac{\alpha_+}{\alpha_-} (p^0-p^3) - \frac{\beta_+}{\alpha_-} (p^1-ip^2)= m,$ 
$\frac{\beta_+}{\beta_-} (p^0+p^3) - \frac{\alpha_+}{\beta_-} (p^1+ip^2)= m,$ 
$\frac{\alpha_-}{\alpha_+} (p^0+p^3) + \frac{\beta_-}{\alpha_+} (p^1-ip^2)= m,$
$\frac{\beta_-}{\beta_+} (p^0-p^3) + \frac{\alpha_-}{\beta_+} (p^1-ip^2)= m.$ 
One notices that for massless solutions  ($m=0$)  $\psi^{(4)}_{(+)}$ 
and $\psi^{(4)}_{(-)}$ 
decouple. 
Taking the above derivation into account Eq.~(\ref{weylErho0}) transforms into
\begin{eqnarray}
\label{weylEp}
f \, \{(p_{05} + i 2S^{56}\,p_{06}) + \frac{1}{2E}\, \{p_{5} + i 2S^{56}\,p_{6}, Ef\}_{-} \}\, \psi^{(6)}
+ \gamma^0 \gamma^5 \, m \, \psi^{(6)}=0.
\end{eqnarray}
For $x^{(5)}$ and $x^{(6)}$ from Eq.~(\ref{f})  
and for the zweibein from 
Eqs.(\ref{fzwei},\ref{f}) and the spin connection from Eq.(\ref{omegas}) one obtains  
\begin{eqnarray}
\label{weylErho}
if \, \{ e^{i \phi 2S^{56}}\, [\frac{\partial}{\partial \rho} + \frac{i\, 2 S^{56}}{\rho} \, 
(\frac{\partial}{\partial \phi}) -  \frac{1}{2 \,f} \, \frac{\partial f}{\partial \rho }\, 
(1- 2F \, 2S^{56})\,]
\, \} \, \psi^{(6)}
+ \gamma^0 \gamma^5 \, m \, \psi^{(6)}=0.
\end{eqnarray}
Having the rotational symmetry around the axis perpendicular to the plane of the fifth and the sixth 
dimension we require that $\psi^{(6)}$ is the eigen function of the total angular momentum
operator $M^{56}= x^5 p^6-x^6 p^5  + S^{56}= -i \frac{\partial}{\partial \phi} + S^{56}$
\begin{eqnarray}
M^{56}\psi^{(6)}= (n+\frac{1}{2})\,\psi^{(6)}.
\label{mabx}
\end{eqnarray}
Accordingly we write
\begin{eqnarray}
\psi^{(6)}= {\cal N}\, ({\cal A}_{n}\, \stackrel{56}{(+)}\, \psi^{(4)}_{(+)}  
+ {\cal B}_{n+1}\, e^{i \phi}\, \stackrel{56}{[-]}\, \psi^{(4)}_{(-)})\, e^{in \phi}.
\label{mabpsi}
\end{eqnarray}
After taking into account that $S^{56} \stackrel{56}{(+)}= \frac{1}{2} \stackrel{56}{(+)}$, while 
$S^{56} \stackrel{56}{[-]}= -\frac{1}{2} \stackrel{56}{[-]}$ we 
end up with the equations of motion
 for $\A_n$ and $\B_{n+1}$ as follows  
\begin{eqnarray}
&&-if \,\{ \,(\frac{\partial}{\partial \rho} + \frac{n+1}{\rho})  -   
  \frac{1}{2\, f} \, \frac{\partial f}{\partial \rho}\, (1+ 2F)\}  \B_{n+1} + m \A_n = 0,  
\nonumber\\
&&-if \,\{ \,(\frac{\partial}{\partial \rho} - \quad \frac{n}{\rho}) -   
  \frac{1}{2\, f} \, \frac{\partial f}{\partial \rho}\, (1- 2F)\}  \A_{n} + m \B_{n+1} = 0.
\label{equationm56gen1}
\end{eqnarray}
Let us treat first the massless case ($m=0$). Taking into account that $F\frac{f-1}{f \rho} = 
\frac{\partial}{\partial \rho} \ln f^{\frac{F}{2}}$ and that $E=f^{-2}$, it  follows  
\begin{eqnarray}
\frac{\partial \, \ln (\B_n \, \rho^n \,f^{-F -1/2})}{\partial \rho}&=&0,\nonumber\\
\frac{\partial \, \ln (\A_n \, \rho^{-n} \,f^{F -1/2})}{\partial \rho}&=&0.
\label{masslesseq}
\end{eqnarray}
We get correspondingly the solutions
\begin{eqnarray}
\B_n \, e^{in \phi}&=& \B_0 \, e^{in \phi}\, \rho^{-n} f^{F+1/2}, \nonumber\\
\A_n \, e^{in \phi}&=& \A_0 \, e^{in \phi}\, \,\rho^{n} f^{-F+1/2}. 
\label{masslesseqsol}
\end{eqnarray}
Requiring that only normalizable (square integrable) solutions are acceptable 
\begin{eqnarray}
2\pi \, \int^{\infty}_{0} \,E\, \rho d\rho \A^{\star}_{n} \A_{n} && < \infty, \nonumber\\
2\pi \, \int^{\infty}_{0} \,E\, \rho d\rho \B^{\star}_{n} \B_{n} && < \infty, 
\label{masslesseqsolf}
\end{eqnarray}
it follows 
\begin{eqnarray}
&&{\rm for}\; \A_{n}: -1 < n < 2F, \nonumber\\
&&{\rm for}\; \B_{n}: 2F < n < 1, \quad n \;\; {\rm is \;\; an \;\;integer}.
\label{masslesseqsolf1}
\end{eqnarray}
One immediately sees that for $F=0$ there is no solution for the zweibein from Eq.~(\ref{f}). 

Eq.~(\ref{masslesseqsolf1}) tells us that the strength $F$ of the spin connection field 
$\omega_{56 \sigma}$ can make a choice between the two massless solutions $\A_n$ and $\B_n$: 
For 
\begin{eqnarray}
0< 2F \le 1
\label{Fformassless}
\end{eqnarray}
 the only massless solution is the left handed spinor with respect to 
$(1+3)$
\begin{eqnarray}
\psi^{(6)m=0}_{\frac{1}{2}} ={\cal N}_0  \; f^{-F+1/2} 
\stackrel{56}{(+)}\psi^{(4)}_{(+)}.
\label{Massless}
\end{eqnarray} 
It is the eigen function  of $M^{56}$ with the eigenvalue $1/2$. 
No right handed massless 
solution is allowed. 
For the  particular choice  $2F=1$ the spin connection field $-S^{56} \omega_{56\sigma}$ 
compensates the term $\frac{1}{2Ef} \{p_{\sigma}, Ef \}_- $ and the  left handed spinor
with respect to $d=(1+3)$ becomes a constant with respect to $\rho $ and $\phi$.

For $2F=1$ it is easy to find also all the massive solutions of Eq.~(\ref{equationm56gen1}).
%
Introducing $u=\frac{\rho}{2\rho_0}$ and assuming that $2F=1$ one finds from Eq.~(\ref{equationm56gen1}) 
%
\begin{eqnarray}
&&\B_{n+1} = \frac{i}{2 \rho_0 m} \, (1+u^2)\, (\frac{d}{du} - \frac{n}{u} )\,\A^{m}_{n},\nonumber\\ 
&&\{(\frac{1+u^2}{2})^2\, \left( \frac{d^2}{du^2} + \frac{1}{u}\, \frac{d}{du} - \frac{n^2}{u^2}\right)\, +
(\rho_0 \, m)^2
\} \A^{m}_{n} = 0\,.
\label{equationm56u}
\end{eqnarray}
If one expresses $(\frac{\rho}{2\rho_0})^2=  \frac{1-x}{1+x}$, with  $-1 \le \,x\,\le 1 $  for 
$0 \le \rho \le \infty$,  it follows that $f=\frac{2}{1+x}$, $\frac{dx}{du}= \frac{-4u}{(1+u^2)^2}$ and 
$\frac{4\,u^2}{(1+u^2)^2}= (1-x^2)$. Then Eq.~(\ref{equationm56u})  transforms into the equations of motion 
for the associate Legendre polynomials $\A^{(\rho_0 m)^2=l(l+1)}_n = P^{l}_n$, 
if we assume that $(\rho_0\, m)^2 =l(l+1)$  
\begin{eqnarray}
&&\left((1-x^2)\,\frac{d^2}{dx^2} -2x\, \frac{d}{dx} - \frac{n^2}{1-x^2} +  l(l+1) \,\right) \, 
\A^{(\rho_0 m)^2=l(l+1)}_n = 0\,,  
\nonumber\\
&& l(l+1)= (\rho_0 \, m)^2\,, \nonumber\\
&& \B^{(\rho_0 m)^2=l(l+1)}_{n+1} = \frac{-i}{\rho_0 m} \,\sqrt{1-x^2}\, 
\left(\frac{d}{dx} + \frac{n}{1-x^2} \right)\, \A^{(\rho_0 m)^2=l(l+1)}_n \,.
\label{equationm56ux}
\end{eqnarray}
From the above  equations 
we see that for $m=0$, that is for the massless 
case, the only solution 
with $n=0$ exists, which is $\A^{(\rho_0 m)^2=0}_0$, which is a constant (in  
agreement with our discussions  above). 

It is not difficult to prove that there is no normalizable solutions of 
Eq.~(\ref{equationm56ux}) for an arbitrary  $m \,\rho_0 $, which is not of the kind 
$(m \rho_0)^2=l(l+1),$ with  
$l$ an integer and also not for a noninteger $n$.  
The solutions of Eq.~(\ref{equationm56ux}) are, namely, not square 
integrable on the interval $-1 \le x \le 1$ for $l \ne {\rm an \; integer}\,,$ and 
$ \,  \nu \ne  {\rm an \;integer}$. 
$P^{n}_{\nu} (x \to -1+0)$ are unbounded,  going to $\infty$, while they are bounded at 
$(x \to 1-0)$.  One also finds that
$P^{\mu}_{n} \to \infty$, if $ (x \to 1-0)$, unless $\mu= \pm m,\,$ with $ m {\,\rm an\; integer}$. 
(See ref.~\cite{SF}, sect. 5.18, pages 255-258.)

Accordingly  the massive solutions  
with the masses equal to $m  = l (l+1)/\rho_0$ (we use the units in which $c=1=\hbar$) 
and the eigenvalues of $M^{56}$ 
((Eq.~\ref{mabx}))---which is 
the charge as we  see in section~\ref{properties1+3}---equal to $(\frac{1}{2}+n)$, 
with $-l \le n \le l$, $l=1,2,..$, are
\begin{eqnarray}
&&\psi^{(6) (\rho_0 m)^{2}=l(l+1)}_{n+1/2} = \nonumber\\
&&{\cal N}^{l}_{n+1/2} \, \left(
\stackrel{56}{(+)} \psi^{(4)}_{(+)} + 
\frac{i}{2 \sqrt{l(l+1)}} \, \stackrel{56}{[-]} \, \psi^{(4)}_{(-)} \, e^{i \phi} \, (1+u^2)\,
(\frac{d}{du} \, -\frac{n}{u})\, \right)
 {}\cdot
e^{i n \phi} \, \A^{(\rho_0 m)^{2}=l(l+1)}_n 
\,,\nonumber\\
\label{knsol}
\end{eqnarray}
with $\A^{(\rho_0 m)^2=l(l+1)}_n (x)$, which are the associate Legendre polynomials $P^{l}_{n}(x)$, 
where $x= \frac{1-u^2}{1+u^2}$, and $u= \frac{\rho}{2 \rho_0}$~\footnote{
	Rewriting the mass operator $\hat{m}= \gamma^0 \gamma^s f^{\sigma}{}_{s} (p_{\sigma} - 
S^{56} \omega_{56 \sigma} + \frac{1}{2Ef} \{p_{\sigma}, Ef\}_-)$ as a function of 
$\vartheta $ and $\phi$:   
%
$\rho_0 \hat{m}= i \gamma^0\, \{\stackrel{56}{(+)} e^{-i\phi} 
(\frac{\partial}{\partial \vartheta} \, -\frac{i}{\sin \vartheta}
\frac{\partial}{\partial \phi } \, - \frac{1-\cos \vartheta}{\sin \vartheta}) +
\stackrel{56}{(-)} e^{i\phi} 
(\frac{\partial}{\partial \vartheta} \, + \frac{i}{\sin \vartheta}
\frac{\partial}{\partial \phi } ) \},$ 
one can easily show that when  applying $\rho_0 \hat{m}$ and $M^{56}$ 
on $\psi^{(6)\hat{m}^2=l(l+1)}_{n+1/2}$, for $l=1,2,\cdot$,  
one obtains
from Eq.~(\ref{knsol})
$\rho_0 \hat{m}\, \psi^{(6)\hat{m}^2=l(l+1)}_{n+1/2} =
l(l+1) \psi^{(6)\hat{m}^2=l(l+1)}_{n+1/2}, \;\; 
M^{56}\, \psi^{(6)\hat{m}^2= 
l(l+1)}_{n+1/2} =
(n+1/2) \psi^{(6)\hat{m}^2=l(l+1)}_{n+1/2}$, $l=1,2,\cdot$.
A  wave packet,  which is the eigen function of $M^{56}$ with the eigenvalue 
$1/2$,  for example,  can be written as
%
$\psi^{(6)}_{1/2} = \,\sum_{k=0,  \infty} C_{1/2}^k \;\,
{\cal N}_{1/2} \{ 
\stackrel{56}{(+)} \psi^{(4)}_{(+)} + (1 - \delta^{k}_0)
\frac{i}{\sqrt{k(k+1)}}\, \stackrel{56}{[-]} \psi^{(4)}_{(-)}\,e^{i\phi}
\frac{\partial}{\partial \vartheta} \}
Y^{k}_{0}. $
%
The expectation value of the mass operator $ \hat{m}$ on such a wave packet is 
$\sum_{k=0,  \infty} C_{1/2}^{k*} C_{1/2}^{k} \sqrt{k(k+1)}/\rho_0$. 
}.
%
It is not difficult to see that the solutions of Eq.~(\ref{equationm56gen1}) for $2F=1$,
 $\psi^{(6)m=0}_{\frac{1}{2}} $ and 
$\psi^{(6) (\rho_0 m)^2=l(l+1)}_{n+1/2} $, are  normalizable on the infinite disc curved 
into almost $S^2$ ($2 \,\pi\, \int\,\rho d \rho E \,\psi^{(6)(\rho_0 m)^2=l(l+1) \dagger}_{n+1/2} \, 
\psi^{(6)(\rho_0 m)^2=l(l+1)}_{n+1/2} < \infty$, with $E= f^{-2}$). One can show as well 
that 
the eigenstates,  
with the discrete eigenvalues $(\rho_0 m)^2=l(l+1)$, 
are orthogonal   
 ($\int\,d^{2}x E \,
(\psi^{(6)(\rho_0m)^2=l'(l'+1) \dagger}_{n'+1/2} \, 
\psi^{(6)(\rho_0 m)^2=l(l+1)}_{n+1/2})= \delta^{l l'} \delta^{n n'} 
\propto \int d^2 x \, e^{-i(n'-n)\phi}\,\{{\cal B}^{l' +}_{n'+1} \, {\cal B}^{l}_{n+1} 
+ {\cal A}^{l' +}_{n'} \,  {\cal A}^{l}_{n} \} $) for all pairs of $(l,n), (l',n')$, the spectrum 
is obviously discrete as it should be 
for the Hermitean Hamiltonian with the boudary conditions determined by normalizability of states. 

To find  solutions for all $F$ in the interval  $0 < F \le \frac{1}{2}$, besides the 
massless one $\psi^{(6)m=0}_{\frac{1}{2}} $, is  a more tough work.  
Yet one can expect that on the space of normalizable functions the Hamiltonian will stay Hermitean and since 
an infinitesimal change of the constant $F$ from $F=\frac{1}{2}$ to  a tiny smaller $F$ 
can not spoil the discreteness of the 
Hamiltonian eigenvalues, the spectrum would stay discrete. One can see that the current in 
the radial direction is zero for any $F$. We studied these solutions and found the discrete spectrum, 
a paper is in preparation.

(Let us recognize that $e^{i n \phi} \, P^{l}_{n}$ are spherical harmonics $Y^{l}_{n}$. 
Expressing $\rho$ with $\vartheta$, $\frac{\rho}{ 2 \rho_0}= \, 
\sqrt{\frac{1- \cos \vartheta}{1+ \cos \vartheta}}$ we rewrite the equations of 
motion~(Eq.\ref{equationm56gen1})as follows
\begin{eqnarray}
&&(\frac{\partial}{\partial \vartheta} +  \frac{n+1 -(F+1/2)(1-\cos \vartheta)}{\sin \vartheta} )\, 
\B_{n+1} + i \rho_0 \,m \A_n = 0,  
\nonumber\\
&&(\frac{\partial}{\partial \vartheta} + \; \frac{-n +(F-1/2)\,(1-\cos \vartheta)}{\sin \vartheta} ) \,\;
\A_{n} + i \rho_0 \,m \B_{n+1} = 0\,.)
\label{equationm56theta}
\end{eqnarray}
\section{Singularities on an almost $S^2$ sphere} 
\label{singularitiesgaugetransformations}  

In this section we comment on singularities "felt" by a spinor  if a noncompact disc with the zweibein 
from Eq.~(\ref{fzwei}) and the  spin connections from Eq.~(\ref{omegas}) is 
understood as the $S^2$ sphere with a hole on the southern pole.

Intuitively it is not difficult to see that we are in troubles if we want  
the chiral fermion field of Eq.~(\ref{Massless}), that is  
$\psi^{(6)m=0}_{\frac{1}{2}} ={\cal N}_0  \; f^{-F+1/2} 
\stackrel{56}{(+)}\psi^{(4)}_{(+)}$,  on a two dimensional space  
to be an eigenstate of some rotational operator $M^{56}$, if the two dimensional space 
has to have the topology of $S^2$, while the spin of the fermion contributes to $M^{56}$ in 
the "usual way" 
\begin{eqnarray}
M^{56} &=& S^{56} + K^{56},
\label{m56k}
\end{eqnarray}
where $K^{56} $ is the Killing vector, like in Eq.~(\ref{mabx}) ($K^{56}= x^5 p^6-x^6 p^5 $).
Near the starting point (the origin, the northern pole of $S^2$) on the topologically $S^2$ sphere 
the Killing operator functions as the orbital angular momentum ($L^{56}=x^{(5)} p^{(6)}-
x^{(6)} p^{(5)}$) and has to be added to the spin 
part $S^{56}$, just as it is in the flat two-dimensional space. Going away from the starting point 
the action of $M^{56}$ may be more complicated as just a simple sum in Eq.~(\ref{m56k}). 
Because of the $S^{2}$ topology there has to be namely yet another point at which the orbital 
Killing generator eigenvalue goes to zero, since there has to be a point, the south pole, 
which is left invariant under the orbital Killing transportation as it is at the starting point, at the 
north pole. 

It is also easy to see that on the two-dimensional $S^2$, the orientation of the Killing 
transportation in the infinitesimal neighbourhood of this second stable point, the south pole, 
 is in the {\it opposite}  direction with respect to the  orientation of the Killing transportation 
 around the north pole. 
 
If we want to have on $S^2$ only a spinor of  one handedness, let say the spinor 
$\psi^{(6)m=0}_{\frac{1}{2}}$ of Eq.~(\ref{Massless}), then we should   
count at the south pole the orbital  symmetry generator with the opposite sign relative to $S^{56}$  
as we do at the starting point (see Eqs.~(\ref{m56spsi6sp},\ref{spinorsouth})). In order to be  able to 
have on the two-dimensional $S^2$ surface a spinor of only one handedness, we have to let 
the phase rotation generated by $S^{56 }$ part of $M^{56}$ relative to the Killing part at 
the south pole to be of the opposite sign with respect to the north pole. Namely, when we 
consider smaller  and smaller circles around the south pole, the phase of the single handedness 
spinor state must be rotated under $M^{56}$ so that when extrapolating to the south pole the phase 
rotation correspond to the spin, which is inverted relative to the orientation of the two-dimensional 
space of the $S^2$ surface.

Therefore, embedding the $S^2$ sphere into a three-dimensional Euclidean space, it is not surprising that 
if we want a spinor of one handedness and succeed to implement it at the north pole in an outward normal 
direction,  we can hardly implement it at the south pole. We might hope for the compensation by the orbital 
part of $M^{56}$, except at the poles. This means that we could have a state of a handed spinor if 
the wave function goes to zero at at least one of the poles, say the 
southern pole (see Eqs.~(\ref{Massless},\ref{masslesseqsolf1})).

\subsection{Formal introduction of a singular point}
\label{forsingpo}

We might formally introduce at the south pole a special singularity, so that we require the wave function 
instead to behave at the south pole in the usual differentiable way, to be differentiable only after 
being multiplied (corrected) by a phase factor: Instead of $\psi$ we require that 
$e^{i \phi^{\SP}} \, \psi$ is our wave differentiable function in the neighbourhood of the singular point 
at the south pole, the phase factor $e^{i \phi^{\SP}}$ itself behaving singularly.
By making this modified requirement of the differentiability we effectively change the 
orbital angular momentum of the wave function  by one unit of $\hbar$ before we require the wave function 
to be smooth or differentiable. Thereby we have made  the requirement that the actual wave function should 
have a rather unphysical extra bit of a negative angular momentum around the south pole.
We must admit that it looks rather strange from the physical point of view, unless we 
recognize that this smoothness condition is to simulate the non-compactness of the $S^2$ space, which 
only after adding a singular point becomes an $S^2$ at all.

When changing the differentiability of the wave function in the neighbourhood of the singular point 
with the requirement that the wave function must be multiplied by a phase, we recognize that such a phase 
multiplication of the wave function appears when transforming the coordinate system from the northern to 
the southern pole, as we can see in equation (\ref{s}) bellow. This phase transformation of the
wave function requires the appearance of the spin connection field, as can be seen in Eq.(\ref{omegasp}): 
The gauge transformation of any spin connection field (when transforming the coordinate system),   
appears even if the spin connection field is zero and manifests in the second term of this equation.

\subsection{Gauge transformations from the northern to the southern pole}
\label{gaugetrans}

To demonstrate further what does the hole  do in the noncompact space of an almost $S^2$ sphere    
let us transform the coordinate system from the northern to the southern 
pole of the sphere $S^2$ as the $S^2$ would be a sphere made out of an infinite plane 
with the zweibein of a sphere and look at how do 
the equations of motion and the wave functions 
transform correspondingly and how do they demonstrate  the noncompactness of our space.

\begin{figure}
\centering
\includegraphics{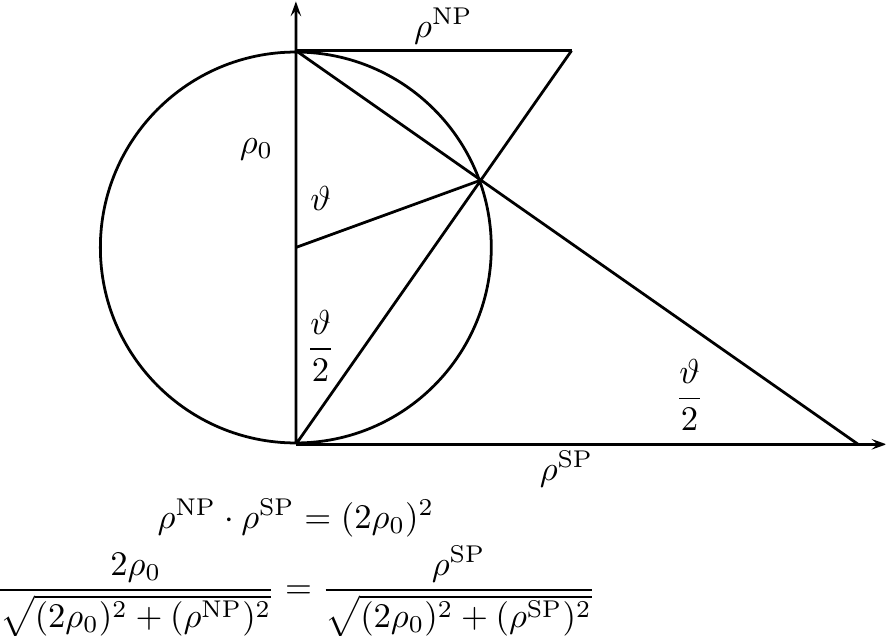}
\caption{Transforming coordinates from the north to the south pole on $S^2$. 
\label{northsouthpole}}
\end{figure}

From Fig.~\ref{northsouthpole} we read
\begin{eqnarray}
\label{xsp}
x^{\NP(5)}&=& (\frac{2\rho_0}{\rho^{\SP}})^2\,x^{\SP(5)},\quad
x^{\NP(6)} = -(\frac{2\rho_0}{\rho^{\SP}})^2\,x^{\SP(6)},
\end{eqnarray}
and
\begin{eqnarray}
\rho^{\SP}\rho^{\NP}&=& (2\rho_0)^2,\quad E^{\NP} \, d^2 x^{\NP} =
E^{\SP} \, d^2 x^{\SP}, 
\end{eqnarray}
where $x^{\NP\sigma }, \sigma=(5),(6)$ stay for up to now used  $x^{\sigma}, \sigma=(5),(6)$, 
while $x^{\SP \sigma }, \sigma=(5),(6)$ stay for coordinates when we put our coordinate system 
at the southern pole and $\rho_0$ is the radius of $S^2$ as before. 
We have $E^{\SP}=(1+ (\frac{\rho^{\SP}}{2 \rho_0})^2)^{-2}$  
and $E^{\NP}=(1+ (\frac{\rho^{\NP}}{2 \rho_0})^2)^{-2}= (\frac{2 \rho_0}{\rho^{\SP}})^4 \,E^{\SP}$.   
We also can write 
$x^{\NP\sigma}=  (\frac{2\rho_0}{\rho^{\SP}})^2\,  (-)^{1+\sigma} \,x^{\SP \sigma}$.

We ought to transform the Lagrange density (Eq.(\ref{weylE})) 
expressed with respect to the coordinates at the northern pole 
\begin{eqnarray}
{\cal L}^{\NP}_{W}&=&\psi^{\NP \dagger}E^{\NP}  \gamma^0 \gamma^s \, ( f^{\NP \sigma}{}_s \, p^{\NP}_{0 \sigma} 
+  \frac{1}{2 E^{\NP} } \, \{p^{\NP}_{\sigma}, E^{\NP} \, f^{\NP \sigma}{}_{s}\}_- )\,\psi^{\NP}, \nonumber\\
p^{\NP}_{0 \sigma} &=& p^{\NP}_{\sigma}- \frac{1}{2} S^{st}\, \omega^{\NP}_{st \sigma},\nonumber\\
f^{\NP \sigma}{}_{s}\, \omega^{\NP}_{s't' \sigma} &=& 
\frac{iF \delta^{\sigma}_{s}\, \varepsilon_{s' t'} x^{\NP}_{\sigma}}{\rho^{2}_{0}}  
\label{weylEnp}
\end{eqnarray}
to the corresponding Lagrange density ${\cal L}^{\SP}_{W}$
expressed with respect to the coordinates at the southern pole by assuming 
\begin{eqnarray}
\psi^{\NP} &=& S \, \psi^{\SP}. 
\label{psinpsp}
\end{eqnarray}
We use the antisymmetric tensor $\varepsilon^{(5)(6)}=1= -\varepsilon^{(5)}{}_{(6)}$.
We recognize that 
\begin{eqnarray}
\label{unpsp}
f^{\NP \sigma}{}_{s}&=&
f^{\SP \sigma'}{}_{t} \;\frac{\partial x^{\NP \sigma}}{\partial x^{\SP \sigma'}}
\; O^{-1 t}{}_{s},\nonumber\\
 f^{\SP \sigma}{}_{s} &=& f^{\SP} \, \delta^{\sigma}_{s}, \quad f^{\SP}= (1+(\frac{\rho^{\SP}}{2 \rho_0})^2).
\end{eqnarray}
The matrix $O$ takes care that the zweibein  expressed with 
respect to the coordinate system at the southern pole 
is diagonal: $f^{\SP \sigma}{}_{s} = f^{\SP}\, \delta^{\sigma}_{s}$ 
\begin{eqnarray}
O = 
\pmatrix{- \cos(2 \phi +\pi)  & - \sin (2 \phi +\pi) \cr
          \;\;\; \sin(2 \phi +\pi)  & - \cos(2 \phi +\pi)  \cr}. 
\label{o}
\end{eqnarray}
Requiring that 
\begin{eqnarray}
S^{-1}  \gamma^0 \gamma^s S \, O^{-1 t}{}_{s}&=& \gamma^0 \gamma^t,
\label{gammanpsp}
\end{eqnarray}
from where 
it follows that $S^{-1}  S^{st} S O^{-1 s'}{}_{s} O^{-1 t'}{}_{t} = S^{s't'}$, 
and recognizing that $p^{\NP}_{\sigma}= 
\frac{\partial x^{\SP \sigma'}}{\partial x^{\NP \sigma}} \, p^{\SP}_{\sigma'}$,  
with $p^{\SP}_{\sigma}= i \frac{\partial}{\partial x^{\SP \sigma}}$, 
we find that 
$\gamma^s \, f^{\NP \sigma}{}_{s}\, p^{\NP}_{0 \sigma}
(=\gamma^s \, f^{\NP \sigma}{}_{s}\,  (p^{\NP}_{ \sigma}  - \frac{1}{2}   
S^{st}\; \omega^{\NP}_{s t  \sigma}))$  
transforms into 
$\gamma^{s} \,f^{\SP \sigma}{}_{s} \, p^{\SP}_{0 \sigma}$
\begin{eqnarray} 
\gamma^{s} \,f^{\SP \sigma}{}_{s} \, p^{\SP}_{0 \sigma}   
&=&\gamma^{s} \,f^{\SP \sigma}{}_{s} \, \{p^{\SP}_{ \sigma} - 
\nonumber\\
&&\frac{1}{2} \, S^{s't'}\,
i \varepsilon_{s' t'} ( \frac{ F\, x^{\SP}_{\sigma}}{f^{\SP}\,(f^{\SP}-1)\, \rho^{2}_0}  + 
2 i\, \frac{ \varepsilon_{\sigma}{}^{\tau}\,x^{\SP}_{\tau}}{ (2\,\rho_{0})^{2} (f^{\SP}-1)} )\}.
\label{gammafposigma}
\end{eqnarray}
In the above equation we took into account that $\omega^{\NP}_{s t  \sigma }$ transforms into 
$O^{-1 s'}{}_{s}\,O^{-1 t'}{}_{t}\,\frac{\partial x^{\SP \sigma'}}{\partial x^{\NP \sigma}} \,
\,
(\omega^{\NP}_{s' t'  \sigma' } + O_{s' t"} 
(\frac{\partial \quad}{\partial x^{\NP \sigma"}} O^{-1 t"}{}_{t'})
\, \frac{\partial x^{\NP \sigma"}}{\partial x^{\SP \sigma'}})$, from where it follows that 
$\omega^{\NP}_{s t  \sigma }$ transforms into 
\begin{eqnarray}
&& O^{-1 s'}{}_{s}\,O^{-1 t'}{}_{t}\,\frac{\partial x^{\SP \sigma'}}{\partial x^{\NP \sigma}} \,
\omega^{\SP}_{s' t'  \sigma' },\nonumber\\
\omega^{\SP}_{s t  \sigma }&=& i  \varepsilon_{s t} \,\{ \frac{ F \, x^{\SP}_{\sigma}}{
 f^{\SP}\, \rho^{2}_{0}\, (f^{\SP}-1)} +2i \frac{ \varepsilon_{\sigma}{}^{\tau}\, x^{\SP}_{\tau}}{(2\rho_0)^2 \,
 (f^{\SP}-1)}\}.
\label{omegasp}
\end{eqnarray}
Similarly we  transform the term 
$\gamma^s \, \frac{1}{2 E^{\NP} } \, \{p^{\NP}_{\sigma}, E^{\NP}\, 
f^{\NP \sigma}{}_{s}\}_- $ into 
\begin{eqnarray}
 \gamma^s ( \frac{1}{2 E^{\SP} } \,  \{p^{\SP}_{\sigma}, E^{\SP}\, 
f^{\SP \sigma}{}_{s}\}_-  + \frac{1}{2} f^{\SP \sigma}{}_{s} 
\{p^{\SP}_{\sigma}, \ln(\frac{\rho^{\SP}}{2 \rho_0})^2\}_{-}\,).
\end{eqnarray}
The action $\int d^2 x^{\NP}{\cal L}^{\NP}_{W}$, with the 
density from Eq.(\ref{weylE}), transforms, when the coordinate system 
is put at the southern pole, as follows 
\begin{eqnarray} 
\int d^2 x^{\NP} {\cal L}^{\NP}_{W}&=&
\int  d^2 x^{\SP} \psi^{\SP \dagger} E^{\SP}S^{\dagger}   \gamma^0 \gamma^s \,  (f^{\SP \sigma'}{}_t \; 
\frac{\partial x^{\NP \sigma}}{\partial x^{\SP \sigma'}}\, 
O^{-1 t}{}_{s}\, \frac{\partial x^{\SP \sigma"}}{\partial x^{\NP \sigma}}\,
\, p^{\SP}_{0 \sigma"} 
+ \nonumber\\
&& 
\,\frac{1}{2 E^{\SP} } \, \{p^{\SP}_{\sigma}, E^{\SP} \, f^{\SP \sigma}{}_{s}\}_- + \frac{1}{2}\,
 f^{\SP \sigma}{}_{s} 
\{p^{\SP}_{\sigma}, \ln( f^{\SP}-1)\}_{-}\,)\,S \,\psi^{\SP},
\label{actionweylEsp}
\end{eqnarray}
which leads to the Lagrange density
\begin{eqnarray} {\cal L}^{\SP}_{W}&=&\psi^{\SP \dagger}E^{\SP}  \gamma^0 \gamma^s \, ( f^{\SP \sigma}{}_s \, p^{\SP}_{0 \sigma} 
+ \nonumber\\
&& 
\frac{1}{2 E^{\SP} } \, \{p^{\SP}_{\sigma}, E^{\SP} \, f^{\SP \sigma}{}_{s}\}_- + \frac{1}{2}\,
 f^{\SP \sigma}{}_{s} 
\{p^{\SP}_{\sigma}, \ln(\frac{\rho^{\SP}}{2 \rho_0})^2\}_{-}\,)\,\psi^{\SP}.
\label{weylEsp}
\end{eqnarray}
The requirement that $S^{-1} \gamma^0 \gamma^s \,S \,  O^{-1 t}{}_{s} = \gamma^0 \gamma^t$
is fulfilled by the operator 
$S= e^{-i S^{56} \omega_{56}}$, and $\omega_{56}= 2 \phi +\pi$, so that in the space 
of the two vectors $(\stackrel{56}{(+)}\psi^{(4)}_{(+)}, \stackrel{56}{[-]}\psi^{(4)}_{(-)})$ 
\begin{eqnarray}
S = 
\pmatrix{e^{i (\phi^{\NP}+ \frac{ \pi}{2})}  & 0 \cr
          0  & e^{-i (\phi^{\NP}+ \frac{ \pi}{2})}  \cr}, 
\label{s}
\end{eqnarray}
with $\phi^{\NP} = - \phi^{\SP}$, while we have
\begin{eqnarray}
\gamma^0 \gamma^5 = 
\pmatrix{ 0  & -1 \cr
          -1 &  0 \cr},
\gamma^0 \gamma^6 = 
\pmatrix{ 0  & \; i \cr
          -i &  0 \cr}.          
\label{gamma56}
\end{eqnarray}

Let us look  how does an eigenstate of $M^{ab}$ from Eq.~(\ref{mabx}), expressed with respect to 
the coordinate at the northern pole 
\begin{eqnarray}
\psi^{\NP(6)}_{n+\frac{1}{2}}= 
(\alpha_{n}(\rho^{\NP}) \stackrel{56}{(+)} \psi^{(4)}_{(+)} + i \beta_{n}(\rho^{\NP}) \stackrel{56}{[-]}
\psi^{(4)}_{(-)} \; e^{i \phi^{\NP}})\, e^{i n \phi^{\NP}}, 
\label{psi6np}
\end{eqnarray}
with the property 
\begin{eqnarray}
M^{\NP 56} \psi^{\NP(6)}_{n+\frac{1}{2}}= (n+\frac{1}{2})\,\psi^{\NP(6)}_{n+\frac{1}{2}}, 
\label{mab}
\end{eqnarray}
where 
$ M^{\NP 56} = (S^{56} -i \frac{\partial}{\partial \phi^{\NP}})\,$, look like  
when we put the coordinate system at the southern pole. 
When putting the coordinate system at the southern pole not only $\phi^{\NP}$ transforms 
into $-\phi^{\SP}$, but also $\gamma^{6}$  goes into $-\gamma^{6}$, accordingly 
\begin{eqnarray}
\label{spinorsouth}
&&\stackrel{56}{(+)} \quad {\rm goes \;\; into }  \quad \stackrel{56}{(-)}\nonumber\\
&&\stackrel{56}{[-]} \;\quad {\rm goes \;\; \,into }  \quad \stackrel{56}{[+]}, 
\end{eqnarray}
therefore $S^{56}\; \stackrel{56}{(-)}= - \frac{1}{2}\;\stackrel{56}{(-)}  $ and 
$S^{56}\; \stackrel{56}{[+]}=  \frac{1}{2}\;\stackrel{56}{[+]}  $. 
Taking into account Eqs.~(\ref{spinorsouth}, \ref{s}, \ref{o})   we obtain
\begin{eqnarray}
&&\psi^{\SP (6)}_{n+\frac{1}{2}} (x^{\NP \tau})= 
S \;\psi^{\NP(6)}_{n+\frac{1}{2}}(x^{\NP \tau}(x^{\SP \tau})) \nonumber\\
&& =
(i \alpha_{n}(\frac{(2 \rho_0)^2}{\rho^{\SP}})\, e^{-i \phi^{\SP}}\, \stackrel{56}{(-)} \psi^{(4)}_{(+)} + 
 \beta_{n}(\frac{(2 \rho_0)^2}{\rho^{\SP}})\, \stackrel{56}{[+]} \psi^{(4)}_{(-)} )\, e^{-i n \phi^{\SP}}\nonumber\\
 && =
(i \alpha^{\SP}_{-(n+1)} \stackrel{56}{(-)} \psi^{(4)}_{(+)} + 
 \beta^{\SP}_{-n}\,e^{i \phi^{\SP}} \, \stackrel{56}{[+]} \psi^{(4)}_{(-)} )\, e^{-i (n +1)\phi^{\SP}}. 
\label{spsi6sp}
\end{eqnarray}
When evaluating  $M^{\SP 56} = (S^{56} + i \frac{\partial}{\partial \phi^{\SP}})\,$ on 
$S \;\psi^{\NP(6)}_{n+\frac{1}{2}}(x^{\NP \tau}(x^{\SP \tau}))$ it follows

\begin{eqnarray}
(S^{56} + i \frac{\partial}{\partial \phi^{\SP}})\; S 
\;\psi^{\NP(6)}_{n+\frac{1}{2}}(x^{\NP \tau}(x^{\SP \tau})) &=&
(n+\frac{1}{2}) \; S \;\psi^{\NP(6)}_{n+\frac{1}{2}}. 
\label{m56spsi6sp}
\end{eqnarray}

Accordingly the massless state $\psi^{\NP(6)m=0}_{\frac{1}{2}} = {\cal N}^{\NP}_0 \, f^{\NP (-F +\frac{1}{2})}\, 
\stackrel{56}{(+)}\, \psi^{(4)}_{(+)}$ from Eq.~(\ref{Massless})  looks, when transforming  the 
coordinate system from the northern to the southern pole, as 
\begin{eqnarray}
\psi^{\SP(6)m=0}_{\frac{1}{2}} &=& {\cal N}^{\SP}_0\, 
(f^{\SP} \,(\frac{2 \rho_0}{\rho^{\SP}})^2)^{(-F +\frac{1}{2})}\, \stackrel{56}{(-)}\, \psi^{(4)}_{(+)} 
\, e^{-i \phi^{\SP}}.
\label{psisp}
\end{eqnarray}
%

Taking into account that  $x^{\SP(5)} + i 2S^{56} x^{\SP(6)} = \rho^{\SP} \, e^{-i 2S^{56}\phi^{\SP}} $  and  
$\frac{\partial \quad}{\quad \partial x^{\SP(5)}} + i 2S^{56}\, \frac{\partial \quad}{\quad \partial x^{\SP(6)}}
= e^{-i 2 S^{56} \phi^{\SP}}\, (\frac{\partial }{\partial \rho^{\SP}} - i 2 S^{56} 
\frac{1}{\rho^{\SP} }\, \frac{\partial\quad}{\partial \phi^{\SP}})$ 
we can write the equations of motion as
\begin{eqnarray}
\label{weylErhosp}
&&if \,  e^{-i \phi^{\SP} 2S^{56}}\, \{(\frac{\partial\quad}{\partial \rho^{\SP}} - 
\frac{i\, 2 S^{56}}{\rho^{\SP}} \, 
\frac{\partial\quad}{\partial \phi^{\SP}}) +  S^{56}\, \frac{1}{\rho^{\SP}}(
	\frac{4 F}{f^{\SP}} - 2 \cdot  2 S^{56})\nonumber\\
	&&+
	\frac{1}{\rho^{\SP}}\, (1 - \frac{f^{\SP}-1}{f^{\SP}})
	\} \, \psi^{(6)}+ \gamma^0 \gamma^5 \, m \, \psi^{(6)}=0.
\end{eqnarray}
For $\psi^{\SP (6)}_{n+ \frac{1}{2}} = ({\cal A}_{-(n+1)
} e^{-i \phi^{\SP}} \,\stackrel{56}{(-)}\,
\psi^{(4)}_{(+)} 
+ {\cal B}_{-n} \,\stackrel{56}{[+]}\, \psi^{(4)}_{(-)})\, e^{-in\phi^{\SP}}$ we find the equations 
for ${\cal A}_{-(n+1)}$ and ${\cal B}_{-n}$
\begin{eqnarray}
\label{weylErhosprho}
&&-if \, \{(\frac{\partial\quad}{\partial \rho^{\SP}} + \frac{-n}{\rho^{\SP}} \, )  + 
\frac{1}{\rho^{\SP}}(\frac{2 F +1}{f^{\SP}} - 1)\} \, {\cal B}_{-n} + m \, {\cal A}_{-(n+1)} = 0,
\nonumber\\
&&-if \, \{(\frac{\partial\quad}{\partial \rho^{\SP}} + \frac{n+1}{\rho^{\SP}} \, 
)  + \frac{1}{\rho^{\SP}}(
	\frac{-2 F +1}{f^{\SP}} - 1)\} \, {\cal A}_{-(n+1)} + m \, {\cal B}_{-n} = 0.	
\end{eqnarray}
When using $f^{\SP}\frac{\partial\quad}{\partial \rho^{\SP}}= \frac{1}{\rho_0}\, 
\frac{\partial\quad}{\partial \vartheta^{\SP}} $ and $\frac{f^{\SP}}{\rho^{\SP}} = \frac{1}{\rho_0}\,
\frac{1}{\sin \vartheta^{\SP}}$ Eq.(\ref{weylErhosprho}) transforms into 
\begin{eqnarray}
\label{weylErhosptheta}
&&(\frac{\partial\quad}{\partial \vartheta^{\SP}} + 
\frac{-n-1 + (F + \frac{1}{2})(1+\cos \vartheta^{\SP})}{\sin \vartheta^{\SP}})
\, {\cal B}_{-n} + i \rho_0 m \, {\cal A}_{-(n+1)} = 0.\nonumber\\
&&(\frac{\partial \quad}{\partial \vartheta^{\SP}} + 
\frac{n +   (-F + \frac{1}{2})(1+\cos \vartheta^{\SP})}{\sin \vartheta^{\SP}}) 
\, {\cal A}_{-(n+1)} + i\rho_0 m \, {\cal B}_{-n} = 0.	
\end{eqnarray}
Again we find for $2F =1$
\begin{eqnarray}
&&\{\frac{1}{\sin \vartheta} \frac{\partial}{\partial \vartheta}(\sin \vartheta 
\frac{\partial}{\partial \vartheta} ) + [(\rho_0 m)^2 - \frac{n^2}{\sin^2\vartheta}]\} \A_{-(n+1)} =0,\nonumber\\
&&{\cal B}_{-n}= \, i\, \frac{1}{(\rho_0 m)^2} \, (\frac{\partial\quad}{\partial \vartheta^{\SP}}  +  
\frac{n }{\sin \vartheta^{\SP}}) 
\, {\cal A}_{-(n+1)}.
\label{sphthetasp}
\end{eqnarray}

Let us conclude this section by recognizing that we 
 have at the south pole allowed a certain special singularity
which is of the following type:
Around a point in the 2-dimensional space - the singular point - we
let the phase of the wave function rotate so that it turns around $2\pi$
as one goes around $2\pi$ in the direction to the singular point
\textit{i.e.} as $\phi$ goes around. This would for a properly smooth
function only be allowed provided that  the magnitude of the wave function
decreases linearly with the distance to the singular point.
Of course, from the point of view of the structure of the singularity 
we can make a gauge transformation and replace the just mentioned phase 
rotation of the wave function by a singular (essentially constant) value
of the spin connection on the circles around the singular point.

\section{Spinors and the gauge fields in $d=(1+3)$}
\label{properties1+3}
To study how do spinors couple to the Kaluza-Klein gauge fields in the case of 
$M^{(1+5)}$, ``broken'' to 
$M^{(1+3)} \times S^2$ with the radius of $S^2$ equal to  $\rho_0$ and with 
the spin connection field 
$\omega_{st \sigma} = i4F \varepsilon_{st} \frac{x_{\sigma}}{\rho}\frac{f-1}{\rho f}$
we first look for (background) gauge gravitational fields, which preserve the rotational symmetry 
around the axis through the northern and southern pole.
Requiring that the symmetry determined by the Killing vectors of Eq.~(\ref{killings}) 
(following ref.~\cite{hnkk06}) with $f^{\sigma}{}_{s} = f \delta^{\sigma}_{s}, f^{\mu}{}_s=0, 
e^{s}{}_{\sigma}= f^{-1} \delta^{s}_{\sigma}, e^{m}{}_{\sigma}=0,$ is preserved, we find 
for the background vielbein field  
\begin{eqnarray}
e^a{}_{\alpha} = 
\pmatrix{\delta^{m}{}_{\mu}  & e^{m}{}_{\sigma}=0 \cr
 e^{s}{}_{\mu} & e^s{}_{\sigma} \cr},
f^{\alpha}{}_{a} =
\pmatrix{\delta^{\mu}{}_{m}  & f^{\sigma}{}_{m} \cr
0= f^{\mu}{}_{s} & f^{\sigma}{}_{s} \cr},
\label{f6}
\end{eqnarray}
with 
\begin{eqnarray}
\label{background}
f^{\sigma}{}_{m} &=& K^{(56)\sigma} B^{(5)(6)}_{\mu} f^{\mu}{}_{m} = 
\varepsilon^{\sigma}{}_{\tau} x^{\tau} A_{\mu} \delta^{\mu}_{m}, \nonumber\\
e^{s}{}_{\mu} &=& - \varepsilon^{\sigma}{}_{\tau} x^{\tau} A_{\mu} e^{s}{}_{\sigma}, 
\end{eqnarray}
 $ 
s=5,6; \sigma = (5),(6)$.  
Requiring that correspondingly the only nonzero torsion fields are those from 
Eq.~(\ref{T}) 
we find for the spin connection fields 
\begin{eqnarray}
\omega_{st \mu} =  \varepsilon_{st}  A_{\mu},
\quad \omega_{sm \mu} = 
\frac{1}{2}f^{-1}\varepsilon_{s \sigma } x^{\sigma} \delta^{\nu}{}_{m} F_{\mu \nu},
\label{omega6}
\end{eqnarray}
$F_{\mu \nu}= A_{[\nu,\mu]}$. 
 The $U(1)$ gauge field $A_{\mu}$ depends only on $x^{\mu}$.
All the other components of the spin connection fields, except (by the 
Killing symmetry preserved)  $\omega_{st\sigma}$ from Eq.~(\ref{weylE}), are zero, 
since for simplicity we allow no gravity in
$(1+3)$ dimensional space. 
The corresponding nonzero torsion fields ${\cal T}^{a}{}_{bc}$ are presented in 
Eq.~(\ref{T}) and in the expressions following this equation, all the other components are zero.

To determine the current, which couples the spinor to the Kaluza-Klein gauge fields 
$A_{\mu}$, we
analyse (as in the refs.~\cite{hnkk06}) the spinor action (Eq.(~\ref{weylE}))
\begin{eqnarray}
{\cal S} &=& \int \; d^dx  \bar{\psi}^{(6)} E \gamma^a p_{0a} \psi^{(6)} =\nonumber\\
&& \int \, d^dx  \bar{\psi}^{(6)} \gamma^s  p_{s} \psi^{(6)}+ \nonumber \\  
&& \int \, d^dx  \bar{\psi}^{(6)} \gamma^m \delta^{\mu}{}_{m} p_{\mu} \psi^{(6)} + \nonumber\\
&& \int \, d^dx   \bar{\psi}^{(6)} \gamma^m  \delta^{\mu}{}_{m} A_{\mu} 
(\varepsilon^{\sigma}{}_{\tau} x^{\tau}
 p_{\sigma} + S^{56}) \psi^{(6)} + \nonumber\\
&& {\rm \; terms } \propto  x^{\sigma} \,{\rm or } \propto   x^{5}  x^{6}.
\label{spinoractioncurrent}
\end{eqnarray}
 Here $\psi^{(6)}$ is a spinor state  in $d=(1+5)$ after the break of $M^{1+5}$ 
 into $M^{1+3} \times $ $S^2$.
 $E$ is for $f^{\alpha}{}_{a}$ from Eq.~(\ref{f6}) equal to $f^{-2}$. 
The  term in the second row in Eq.~(\ref{spinoractioncurrent}) is the mass term  
(equal to zero for the massless spinor), the term in the third row is the kinetic term, 
together with the term in the fourth row  defines  
the  covariant derivative $p_{0 \mu}$ in $d=(1+3)$.  
The terms in the last row  contribute nothing when the integration over 
the disk (curved into a sphere $S^2$) is performed, since they all 
are proportional to $x^{\sigma}$ or to $ \varepsilon_{\sigma \tau} x^{\sigma} x^{\tau}\;$ 
($-\gamma^{m} \,\frac{1}{2}S^{sm} \omega_{s m n} = -\gamma^{m}\,\frac{1}{2}\,f^{-1}
F_{m n}  \varepsilon_{s \sigma} x^{\sigma}$ and $-\gamma^m \,f^{\sigma}{}_{m}\frac{1}{2}
\,S^{st} \omega_{st \sigma}= 
\gamma^m A_m   x^{5}x^{6} S^{st} \varepsilon_{s t} \frac{4iF(f-1)}{f \rho^2}$).

We end up with the current in $(1+3)$
\begin{eqnarray}
j^{\mu} = \int \;E  d^2x \bar{\psi}^{(6)} \gamma^m \delta^{\mu}{}_{m} M^{56}  \psi^{(6)}.
\label{currentdisk}
\end{eqnarray}
 The charge in $d=(1+3)$ is  proportional to the total 
angular momentum  $M^{56} =L^{56} + S^{56}$ around the axis from the southern to the 
northern  pole of $S^2$, but since for the choice of 
$  2 F =1$ (and for any $0 < 2F \le 1 $) in Eq.~(\ref{masslesseqsolf1}) only a left 
handed massless spinor exists,  
with the angular momentum zero, the charge of a massless 
spinor in $d=(1+3)$ is equal to  $1/2$.

The Riemann scalar is for 
the vielbein of Eq.~(\ref{f6}) equal to 
${\cal R}= -\frac{1}{2} \rho^2 f^{-2} F^{mn}F_{mn}$.

If we integrate the Riemann scalar 
over the fifth and the sixth dimension, we get $-\frac{8\pi}{3} (\rho_0)^4 F^{mn}F_{mn}$.

\section{Conclusions}
\label{conclusion}

We prove in this paper that  one can  escape from the "no-go theorem" of Witten~\cite{witten}, 
that is one can guarantee the masslessness of spinors and their chiral coupling  to the 
Kaluza-Klein[like] gauge fields when breaking the symmetry from  $d$-dimensional one to 
$M^{(1+3)} \times M^{d-4}$ space, in cases which we call the "effective two dimensionality" 
even without boundaries, as we proposed in the references~\cite{hnkk06}. 
Namely, we can guarantee the above mentioned properties of spinors, 
when the break $M^{(1+3)} \times M^{d-4}$, $d-4> 2$ occurs in a way that vielbeins and 
spin connections are completely flat in all but two dimensions, while the two dimensional 
space, although of finite volume, is noncompact with a particular spin connection contributing 
to the properties of spinors. In our particular case it is the zweibein (the zweibein of the $S^2$ sphere 
with a hole  at the southern pole) on an infinite disc, which guarantees that the 
noncompact space has the finite volume and enables, together with the 
spin connection field on this disc ($\omega_{st \sigma}= i\,F \, \varepsilon_{st} 
\frac{x_{\sigma}}{f \, \rho^{2}_{0}}\,$, the $\omega_{st \sigma}$ field breaks the parity symmetry 
and the sign of $F$ makes a choice of the handedness of the massless state) 
that only one normalizable spinor state (of particular handedness) is massless,  
carrying the Kaluza-Klein charge of $\frac{1}{2}$ and 
coupling chirally to the 
corresponding Kaluza-Klein gauge field.  
Let us add that requiring normalizability of  states 
in extra dimensions guarantees that  states are normalizable in the whole $d=(1+ (d-1))$ space. 

Since the spin connection strength  is determined only within an interval  
 ($0 < 2F \le 1 $), what we proposed  is not a fine tuning. 
Taking (in the absence of fermions) the 
action for the gravitational gauge fields with the linear curvature for $d=2$ 
(when any zweibein and any spin connection fulfills the 
corresponding equations of motion), we are allowed to make any choice of a zweibein and  spin connection. 
(This choice leads to nonzero torsion.)

 There is the discrete spectrum of normalizable eigenstates of the Hermitean Hamiltonian on 
 the infinite disc for the chosen 
 zweibein and spin connection of any 
 strength $F$ in the interval ($0 < 2F \le 1 $), as we proved in 
 section~\ref{starting action}.

 The normalizable eigenstates, which are chosen 
 to be at the same time 
the eigenstates of the total angular momentum on the disc $M^{56}= x^{5} p^{6}- x^{6} p^{5} + S^{56}$, 
 with the eigenvalues 
$(n+1/2)$, carry the Kaluza-Klein charge $(n+1/2)$. 
The only massless state carries the charge $(\frac{1}{2})$.
For the choice of $2F=1$ the normalizable massless state is independent of the coordinates on the disc.
The normalizable massive states have the masses 
equal to $k(k+1)/\rho_0, k=1,2,3,..$,
 with $-k \le n \le k$. 
 The spectrum is obviously discrete and stays discrete for 
 all $F$ in the interval $0 < 2F \le 1 $ and for any finite $\rho_0$. 
 The current is for all the solutions and also for all $F$  equal to zero. 
 As long as the Hamiltonian is Hermitean on a disc, fermions can not leave the disc, unless an  
 additional interaction (or a dynamical restoration of the symmetry $M^{(1+5)}$, that is the {\em phase 
 transition}) would force them to go out of the disc, which is not the case for our toy 
 model.

Understanding the infinite disc as a $S^2$ sphere with the southern pole missing, a singularity 
of the type should be recognized:
Around a point in the 2-dimensional space of $S^2$ - the singular point - we
let the phase of the wave function rotate so that it turns around $2\pi$
as one goes around $2\pi$ in the direction to the singular point. 
But from the point of view of the structure of the singularity 
we can make a gauge transformation and replace the just mentioned phase 
rotation of the wave function by a singular (essentially constant) value
of the spin connection on the circles around the singular point.

The possibility  that after the break 
a  two dimensional manifold (with the zweibein of $S^2$, with one point missing and with a particular spin 
connection field) exists allowing only one normalizable massless state 
which is correspondingly mass protested and which couples to the Kaluza-Klein charge, 
opens, to our understanding, 
a new hope for the Kaluza-Klein[like] theories of the elegant version, with only the 
gravity, and 
will help to revive them. 



\appendix*

\section{The Killing vectors and the torsion terms for our model }

The infinitesimal coordinate 
transformations manifesting the symmetry  of $M^{1+3}$ and the $S^2$ are: 
$x^{'\mu}= x^{\mu}, $ $x^{'\sigma}= x^{\sigma} + 
\phi_{A} \,K^{A \sigma}$, with $\phi_{A}$ the parameter of rotations 
 around the axis which goes through both poles and with the 
infinitesimal generators of rotations around this axis $M^{(5)(6)}(= x^{(5)} p^{(6)}- 
x^{(6)} p^{(5)} + S^{(5)(6)})$
\begin{eqnarray}
 K^{A \sigma}= K^{(56) \sigma} = -i M^{(5)(6)} x^{\sigma} =
 \varepsilon^{\sigma}{}_{\tau} x^{\tau}, 
\label{killings}
\end{eqnarray}
with $\varepsilon^{\sigma}{}_{\tau}= -1 = - \varepsilon_{\tau}{}^{\sigma}, \, 
\varepsilon^{(5) (6)}=1. $ The operators $K^{A}_{ \sigma}=f^{-2} 
\varepsilon_{\sigma \tau} x^{\tau}$ fulfil the Killing relation $K^{A}_{ \sigma, \tau} + 
\Gamma^{\sigma'}{}_{\sigma \tau} K^{A}_{ \sigma'} + K^{A}_{ \tau, \sigma} + 
\Gamma^{\sigma'}{}_{\tau \sigma} K^{A}_{ \sigma'}=0,$ (with $\Gamma^{\sigma'}{}_{\sigma \tau}= - 
\frac{1}{2} \, g^{\rho \sigma'} (g_{\tau \rho,\sigma} + g_{\sigma \rho,\tau} - 
g_{\sigma \tau,\rho})$).

From $\gamma^a p_{0a}\gamma^b p_{0b}= p_{0a} p_{0}{}^a - 
i S^{ab} S^{cd}\, {\cal R}_{abcd} + S^{ab}\,{\cal T}^{\beta}{}_{ab} \,p_{0 \beta}$ 
we find for 
the torsion 
\begin{eqnarray}
 {\cal T}^{\beta}{}_{ab}&= &f^{\alpha}{}_{[a} (f^{\beta}{}_{b]})_{, \alpha} + \omega_{[a}{}^{c}{}_{b]} 
f^{\beta}_{c}. 
\label{T}
\end{eqnarray}
 From Eq.~(\ref{T}) we read that to the torsion on $S^2$ both, the zweibein 
 $f^{\sigma}_{\tau}$  and the spin connection $\omega_{st \sigma}$,  
 contribute. While we have on $S^2$ for ${\cal R}_{\sigma \tau} = 
 f^{-2} \eta_{\sigma \tau} \frac{1}{\rho_{0
 }^2} $ and correspondingly for the curvature 
 ${\cal R}= \frac{-2}{(\rho_0)^2}$, we find for the torsion 
 ${\cal T}^{s}{}_{t s'} = {\cal T}^{s}{}_{t \sigma} f^{\sigma}_{s'}$ with 
 %
 ${\cal T}^{5}{}_{ss} =  
 0 = {\cal T}^{6}{}_{ss},\quad s=5,6, $ 
 ${\cal T}^{5}{}_{65} = - {\cal T}^{5}{}_{56} =   -(f_{,6} + \frac{4iF (f-1)}{\rho^2 } x_5), $
 ${\cal T}^{6}{}_{56} = -  
  {\cal T}^{6}{}_{65} = - f_{,5} + \frac{4iF (f-1)}{\rho^2 } x_6.$
  %
  The torsion ${\cal T}^2 = {\cal T}^{s}{}_{t s'} {\cal T}_{s}{}^{t s'} $ 
  is for our particular choice of the zweibein and spin connection fields 
  from Eqs.~(\ref{f},\ref{omegas}) correspondingly   equal to  
  $-\frac{2 \rho^2}{ (\rho_0)^4} (1-  (2F)^2)$.

\end{document}